# A dataset of early blockchain-registered AI agents on Ethereum

Yulin Liu, Quantum Economics AI Lab, Zurich, Switzerland
yulin@quantecon.ai

## Abstract

This study presents a structured dataset of blockchain-registered artificial intelligence agents under the ERC-8004 standard on Ethereum. The dataset integrates on-chain identity records, minting transactions, transfer events, reputation summaries, and individual feedback records, together with resolved off-chain metadata where available. Data were collected from Ethereum mainnet using Web3 RPC queries and processed into tabular form to enable reproducible analysis. The dataset covers 10,000 agents within a defined block range and includes both event-level records and aggregated summaries. It enables empirical research on agent identity formation, reputation systems, service exposure, and early-stage decentralized AI ecosystems. This resource supports studies in blockchain analytics, decentralized trust infrastructure, and the emerging agentic economy.

## Background & Summary

Blockchains[1] have become an important source of large-scale, machine-readable records for studying digital economic activity[2-4]. Prior work has shown that blockchain's immutable provenance and decentralized governance can establish a trustworthy foundation for data exchange. This, in turn, enhances the auditability and integrity of large-scale scientific datasets[5,6]. In the Bitcoin[7-11] context, cohort-based processing has been used to reorganize the full transaction history into economically meaningful datasets and indicators, making a rapidly growing ledger more tractable for research[12,13]. In Ethereum[14-16], data-intensive empirical work has likewise combined on-chain transaction records with protocol-level context to study transaction fee mechanisms and user behavior[17]. Building on this line of research, the present work extends blockchain data curation from transactions and fees to a new emerging object of study: on-chain registered artificial intelligence agents.

Recent advances in large language models, autonomous software agents[18-20], and agent-to-agent interaction frameworks have accelerated interest in an "agentic economy." In such an economy, software agents discover one another, exchange services, coordinate tasks, and engage in financial activity. In practice, once an AI agent begins to hold assets, receive payments, trigger transactions, or participate in decentralized applications, it requires a blockchain-compatible identity and wallet layer. However, communication protocols alone do not solve the full coordination problem. Protocols such as Model Context Protocol[21] and agent-to-agent[22] messaging standards define how agents expose capabilities or exchange messages. They do not, however, establish who an agent is, how it can be addressed across systems, or whether it has accumulated credible evidence of trustworthy behavior.

ERC-8004 addresses this gap by proposing a trust infrastructure for autonomous agents built around three linked components: identity, reputation, and validation[23]. The Identity Registry establishes unique on-chain agent identities as ERC-721 tokens[24]. Each agent is associated with a registry namespace, token identifier, ownership record, and a metadata pointer that can resolve to off-chain descriptive files. The Reputation Registry records historical feedback submitted by counterparties, allowing agents to accumulate performance signals over time through quantitative values and flexible user-defined tags. The architecture also includes a Validation Registry for higher-assurance verification of agent outputs. This component remains immature in current deployments and does not yet provide comparable observable data. Taken together, these registries suggest an infrastructure for machine-readable identity, accountability, and trust in multi-agent environments.

Despite the conceptual importance of this standard, ERC-8004 data remain difficult to analyze in practice. Information is distributed across smart contract storage, emitted events, token metadata URIs, and heterogeneous off-chain JSON documents. The protocol is intentionally minimalist on-chain: only the minimum data needed for interoperability and verification are recorded at the contract layer. Richer descriptive content, service endpoints, and operational details are delegated to off-chain metadata. This design improves flexibility, but it also fragments the observable record. As a result, researchers and builders currently lack a structured public dataset that documents the early formation of ERC-8004 agent identity and reputation at scale.

Here we present a dataset of the first 10,000 ERC-8004 AI agents registered on Ethereum mainnet. Observation begins with agent 0, minted at block 24,339,925 on 2026-01-29 10:31:11 UTC, and extends through block 24,839,925, spanning 500,000 Ethereum blocks. All 10,000 agents were in fact minted within the first 2,298 blocks of this window (blocks 24,339,925–24,342,223, a span of approximately 7.7 hours on 2026-01-29); block 24,839,925 serves as the fixed observation block at which dynamic on-chain state was queried (see Methods). Within this window, we document all agent identifiers from 0 through 9,999. For each registered agent, the core dataset records the mint block, mint timestamp, mint transaction hash, owner wallet, agent wallet where available, token URI, metadata hosting type, and lifecycle status. Across the 10,000 agents, 7,856 are observed in a minted_only state and 2,144 in a metadata_linked state (an agent is classified as metadata_linked when its on-chain token URI is set, and minted_only otherwise; metadata_linked does not imply that the linked off-chain document could be retrieved or parsed — see Technical Validation).

To complement these on-chain identity records, we resolve and normalize off-chain metadata for the subset of agents whose token-linked metadata could be retrieved and parsed successfully. In the current release, structured metadata were resolved for 72 agents (of the 2,144 agents carrying a token URI; the resolution-failure analysis is reported in Technical Validation), including agent names, descriptions, image URLs, active flags, x402 support indicators, and counts of registered services, trust entries, and registrations. We additionally extracted 112 service records from metadata, covering service types such as MCP, A2A, OASF, x402, web, docs, and HTTP-JSON-RPC endpoints, along with optional version, skills, and domains fields. These metadata reveal how ERC-8004 identities can serve as a bridge between NFT-style on-chain registration and richer agent capability descriptions hosted off-chain.

The dataset also captures the early reputation layer of ERC-8004. 980 underlying feedback records were observed from 197 unique client addresses. These feedback records store value and precision fields together with flexible tag dimensions, enabling heterogeneous assessment criteria rather than a single fixed rating scheme. In our observation window, 930 feedback entries contain a first tag and 899 contain a second tag; common examples include liveness, trust, quality, helpfulness, and reachability, with paired operational tags such as liveness-check and oracle-screening. No feedback revocations were observed in this release. This structure is important because it reflects the protocol's design philosophy: trust signals are intended to be composable, application-specific, and extensible rather than reduced to one universal score.

This dataset is designed to support both scientific and applied reuse. For researchers, it provides an early empirical record of how agent identity and reputation are instantiated on-chain, enabling work on trust formation, service discovery, reputation systems, agent ecosystems, and the transaction-level economics of agent minting (e.g., gas usage and mint cost). For builders, the dataset provides a practical index of registered ERC-8004 agents, metadata endpoints, and service declarations that can support dashboards, registries, search tools, analytics pipelines, and protocol monitoring. More broadly, the dataset offers a reproducible baseline for studying the earliest large-scale wave of on-chain AI agent registration on Ethereum. This parallels how earlier blockchain datasets enabled systematic study of transactions, fees, and user activity in earlier stages of crypto-economic development[25-28]. Because the Validation Registry component of ERC-8004 is not yet meaningfully populated on-chain, this release characterizes the identity and reputation layers only and is not intended as a complete record of the full ERC-8004 trust stack.

Relative to existing Ethereum and NFT datasets, this resource is distinguished less by the raw on-chain primitives it records (ERC-721 mints and transfers) than by the off-chain semantic layer it links to each token: agent capability declarations, service endpoints (e.g., MCP, A2A, x402, OASF), and a multi-dimensional, tag-based reputation graph defined by ERC-8004. General NFT or token datasets do not capture machine-readable agent identity, declared interoperability protocols, or peer-submitted reputation feedback, and therefore cannot support questions about how autonomous agents advertise capabilities, how reputation accrues across counterparties, or how an agent-identity layer forms in its first hours of existence. By jointly indexing identity, minting economics, ownership transfers, off-chain metadata, declared services, and reputation feedback under a single agent_id key, the dataset enables such questions to be studied on a common, reproducible substrate.

## Methods

The overall data collection and processing pipeline is illustrated in Fig. 1.

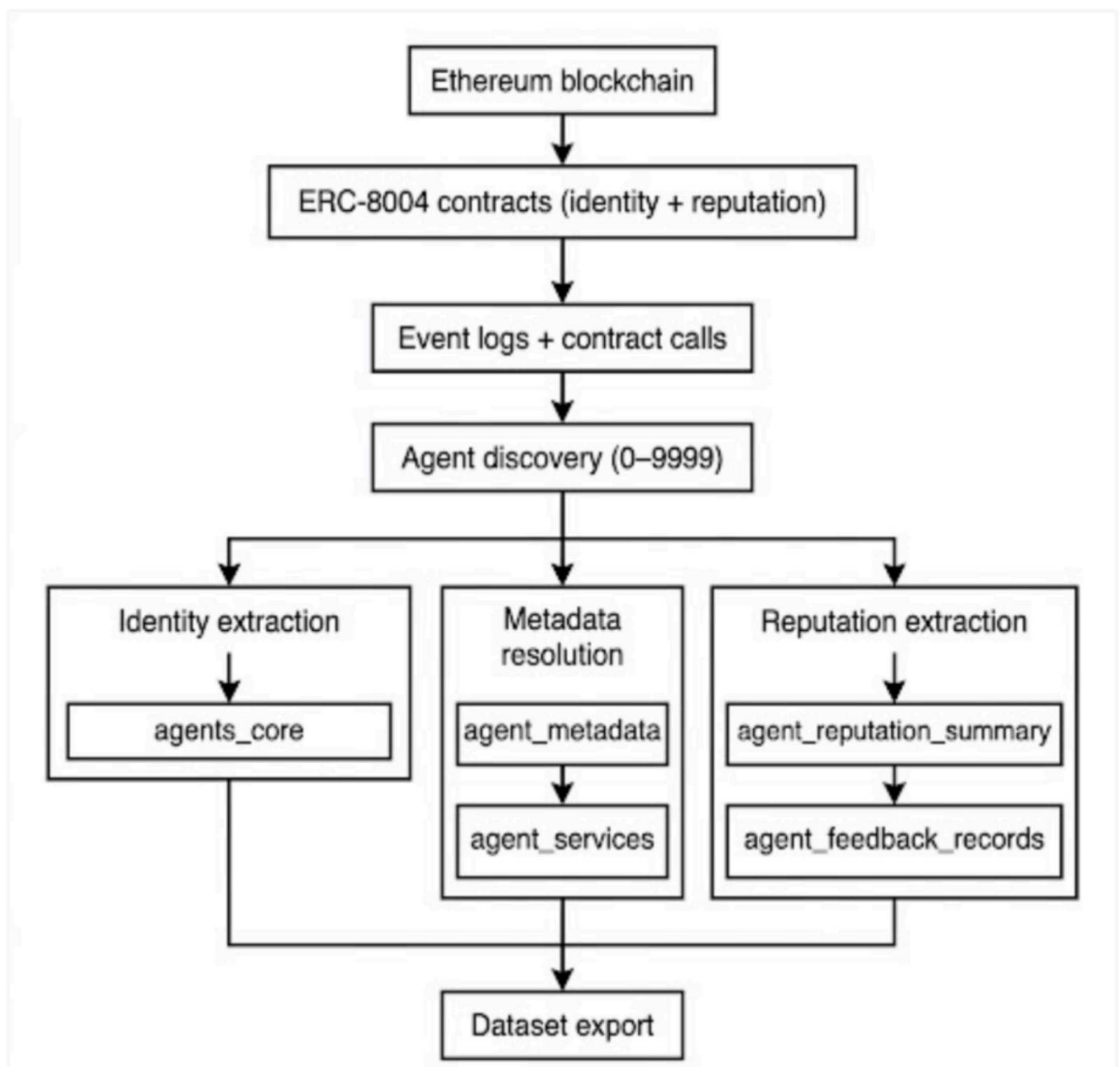

*Figure 1: Data collection and processing pipeline for ERC 8004 agents. On-chain data were read from the ERC-8004 identity and reputation registry contracts on the Ethereum mainnet through an Ethereum archive node (Alchemy JSON-RPC), with all state-dependent calls pinned to a fixed observation block (24,839,925). Agent identities were first discovered by scanning Transfer events and isolating mint events (sender topic = zero address). Each agent was then processed through three parallel stages: (i) identity extraction (ownerOf, tokenURI, mint block and timestamp); (ii) metadata resolution (token URI → HTTP/IPFS → validated JSON, service endpoints, cross-chain registrations); and (iii) reputation extraction (getClients, getSummary, getLastIndex, readFeedback). These stages populate the eight relational tables — agents_core, agent_metadata, agent_services, agent_reputation_summary, agent_feedback_records, mint_economics, transfer_history, and crosschain_registrations — which are exported as CSV for downstream use, plus a derived issuer-screening table and a raw-metadata snapshot.*

### *Data sources and study scope*

Data were collected from the Ethereum mainnet, focusing on the ERC-8004 identity registry at 0x8004A169FB4a3325136EB29fA0ceB6D2e539a432 and the reputation registry at 0x8004BAa17C55a88189AE136b182e5fdA19dE9b63. The dataset corresponds to the namespace eip155:1:0x8004A169FB4a3325136EB29fA0ceB6D2e539a432. Collection was performed using Python, Web3.py, Requests, and a PostgreSQL-compatible database. On-chain state and event logs were obtained from an Ethereum archive node accessed through the Alchemy JSON-RPC API; all state-dependent reads used historical eth_call and eth_getLogs pinned to the fixed observation block. The study covers agents with identifiers 0–9999. The observation window spans blocks 24,339,925 to 24,839,925, anchored at a fixed observation block 24,839,925 for state queries (the fixed observation block is the single block height at which all dynamic contract state — ownership, token URIs, reputation summaries, and feedback — was read, so that every agent is described as of one consistent moment).

Transfer events were reconstructed from on-chain logs emitted by the ERC-8004 identity registry contract using an event-based scanning pipeline. The contract state contains no indexed transfer history. Because full-history log traversal is costly, we extracted events with an optimized scanner: logs were requested in fixed 500-block chunks with adaptive chunk-size reduction on RPC errors, bounded concurrency, request rate-limiting, exponential-backoff retries, a serial second-pass retry for failed ranges, and checkpointing so the scan could be run in successive batches. Using this pipeline, we scanned the first 100,000 blocks of the observation window (blocks 24,339,925–24,439,925). All detected transfer events within this interval were normalized into a structured table, including source and destination addresses, agent identifiers, and associated transaction metadata. Mint operations, which emit transfer-like logs in ERC-8004, were programmatically distinguished and cross-referenced with the mint_economics table to avoid duplication.

### *Snapshot design*

A fixed observation block was used to ensure that all extracted data represent a consistent state of each agent at a specific point in time. Immutable attributes, such as agent identifiers and minting information, are independent of the observation block. Dynamic attributes—ownership, reputation metrics, feedback counts, service declarations, and lifecycle status—depend on the cumulative history of on-chain events and therefore vary across blocks. Querying contract state at a single predefined block ensures consistency, reproducibility, and comparability across agents in downstream use cases.

### *Agent discovery*

Agents were identified by scanning Transfer(address,address,uint256) events emitted by the identity registry. Mint events were detected by filtering logs where the sender topic equals the zero address. Discovered token IDs were filtered to the range 0–9999 and deduplicated.

### *Identity extraction*

For each agent, the identity state was queried at the observation block using ownerOf(tokenId) and tokenURI(tokenId). Mint block and transaction hash were obtained from event logs, and timestamps were derived from block headers. Records were stored in agents_core with normalized addresses and transaction hashes. Metadata hosting type was inferred from URI prefixes. Lifecycle status was assigned as metadata_linked if a token URI was present and minted_only otherwise.

### *Mint transaction data*

Mint transactions were enriched by querying transaction objects, receipts, and block headers. Extracted fields include gas used, gas price, transaction value, nonce, block gas limit, and block gas used. Mint cost was computed as gas_used × gas_price. Results were stored in mint_economics, keyed by transaction hash. A single mint transaction could register multiple agents; mint_economics therefore contains 3,028 unique mint transactions for the 10,000 agents.

### *Metadata resolution*

Token URIs were resolved to HTTP-accessible endpoints. IPFS URIs were mapped to gateway URLs, and base64-encoded data URIs were decoded when present. HTTP responses were required to be non-empty and valid JSON. Parsed fields include name, description, image URL, active status, x402 support, and counts of services, trust entries, and registrations. Service entries were normalized into agent_services, preserving order, endpoint, version, and optional skills and domains. Cross-chain registrations were parsed and stored separately. Agents without resolvable metadata were retained in the dataset. For each successfully retrieved document, the resolved URL and retrieval timestamp were recorded. In the revised release, the resolved off-chain metadata documents were additionally archived in metadata_raw_snapshot.jsonl, with source type, retrieval status, IPFS CID where applicable, and SHA-256 hashes for the archived byte content. Two HTTP documents that returned 404 at archival time are flagged as failed in the snapshot, while their parsed fields remain in agent_metadata.csv.

### *Reputation extraction*

Reputation data were collected at the observation block using the ERC-8004 reputation registry. For each agent, client addresses were obtained via getClients. Aggregate statistics were computed using getSummary. Individual feedback records were retrieved by iterating over indices returned by getLastIndex and reading entries via readFeedback. Feedback records include value, decimals, tags, and revocation status, and were stored in agent_feedback_records. Aggregate results were stored in agent_reputation_summary.

### *Database construction*

The dataset was organized as a relational schema with tables for identity, metadata, services, cross-chain registrations, reputation summaries, feedback records, and mint transaction data. Records were inserted using upsert operations. Child tables were refreshed by deleting and rewriting rows per agent to maintain consistency.

### *Concurrency and fault tolerance*

Data collection used bounded thread pools with staged execution. Remote Procedure Call (RPC) requests were rate-limited using semaphores and global throttling, with exponential backoff and retry for transient errors. A second-pass retry stage reprocessed failed records serially. Metadata requests were executed with explicit timeouts and strict JSON validation.

### *Inclusion and normalization*

All agents in the target ID range were retained regardless of metadata or reputation availability. Addresses and transaction hashes were normalized to lowercase. Timestamps were stored in UTC. Numeric reputation values were stored as raw integers with explicit decimal precision. Revoked feedback entries were preserved. Direct contact endpoints (email addresses) appearing in a small number of off-chain service declarations were removed from the released agent_services table (see Usage Notes); all other declared endpoints were retained.

## Data Record

The dataset is available at Harvard Dataverse at https://doi.org/10.7910/DVN/HJZW8Q and is released under a CC0 1.0 public-domain dedication[29]. It comprises nine CSV files and one JSON-Lines metadata snapshot, plus a data dictionary, related through a common agent_id key (and, for transaction-level tables, the mint transaction hash). The relational structure is illustrated in Fig. 2.

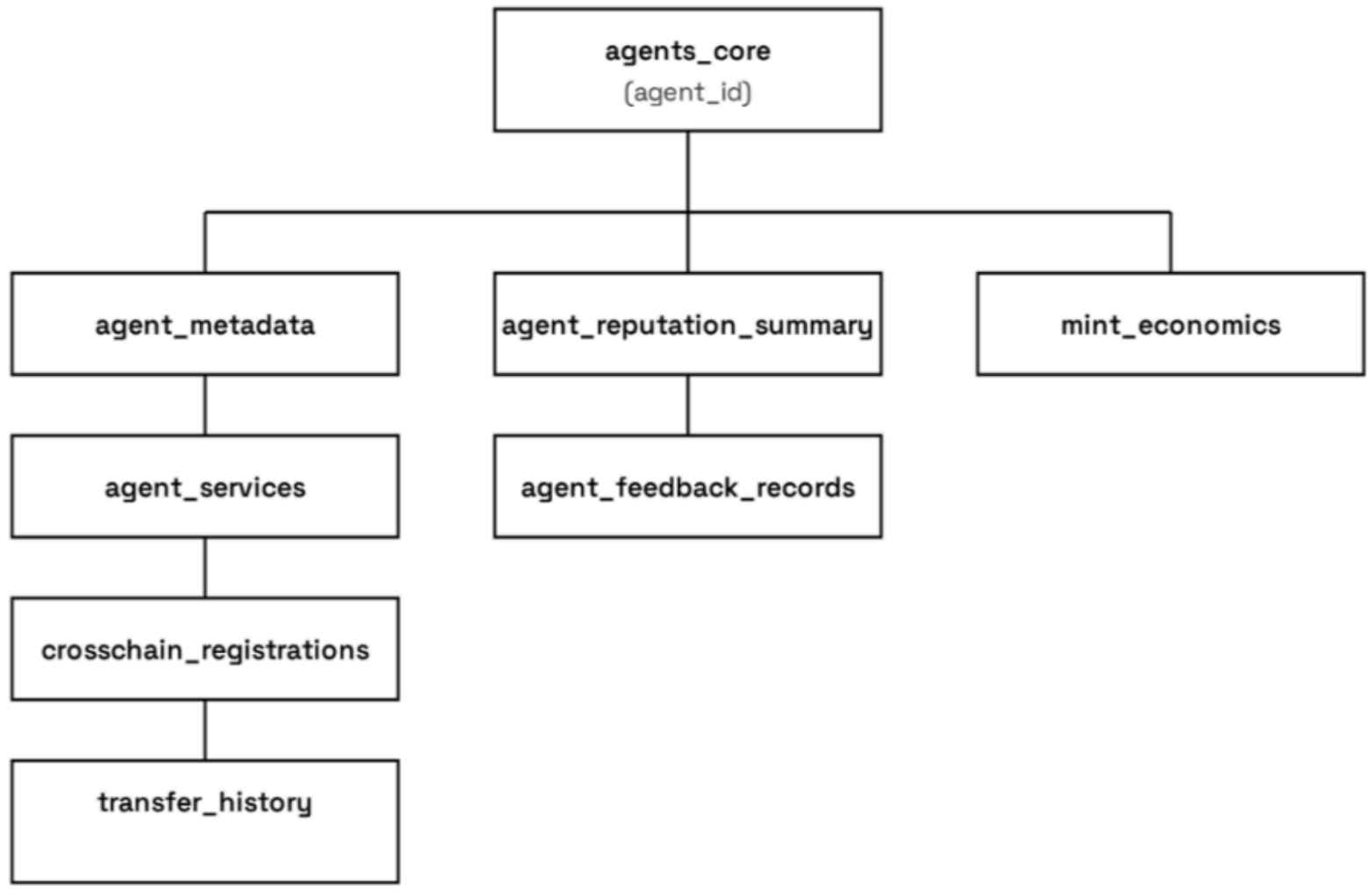

*Figure 2: Relational schema of the ERC 8004 agent dataset. agents_core is the central index (keyed by agent_id) to which the metadata, services, reputation-summary, feedback, transfer, and cross-chain tables join on agent_id, while mint_economics joins on the mint transaction hash—linking on-chain identity, economics, and ownership with off-chain metadata and reputation under common keys.*

**agents_core.csv (10,000 rows):** one row per agent — agent_id, mint_block, mint_timestamp (UTC), mint_tx_hash, owner_wallet, agent_wallet, token_uri, metadata_hosting_type (https/ipfs/other), lifecycle_status (minted_only/metadata_linked).

**agent_metadata.csv (72 rows):** resolved off-chain metadata — metadata_url_resolved, metadata_content_type, metadata_updated_at, name, description, image_url, x402_support, active, service_count, trust_count, registration_count.

**agent_services.csv (112 rows):** declared service endpoints — id, agent_id, service_order, service_name, endpoint, version, skills_json, domains_json.

**agent_reputation_summary.csv (10,000 rows):** per-agent aggregate reputation — agent_id, feedback_count_total, reputation_score_raw, reputation_score_decimals, client_count.

**agent_feedback_records.csv (980 rows):** individual feedback entries — agent_id, client_address, feedback_index, value_raw, value_decimals, tag1, tag2, is_revoked.

**mint_economics.csv (3,028 rows):** per-mint-transaction economics — mint_tx_hash, gas_used, gas_price_wei, mint_cost_eth, tx_value_wei, nonce, block_gas_limit, block_gas_used.

**transfer_history.csv (1,961 rows):** ownership-transfer events — id, agent_id, from_wallet, to_wallet, tx_hash, block_number, timestamp, log_index.

**crosschain_registrations.csv (79 rows):** cross-chain identity mappings — id, source_agent_id, target_chain_namespace, target_chain_id, target_identity_registry, target_agent_id.

**feedback_issuer_summary.csv (197 rows):** per-issuer screening summary for the reputation layer — client_address, issuer_feedback_count, issuer_agent_count, issuer_share_of_all_feedback, is_self_referential (the issuer equals the rated agent's owner for at least one record), and is_high_volume_issuer (the issuer accounts for ≥1% of all feedback records). These are descriptive screening variables that flag concentrated or self-referential feedback patterns; they do not constitute an adjudication of malicious behaviour.

**metadata_raw_snapshot.jsonl (72 records):** a frozen snapshot of the off-chain metadata documents for the 72 metadata-bearing agents, one JSON object per line, with agent_id, resolved_url, source_type (on-chain data URI / IPFS / HTTP), fetched_at (UTC), http_status, a status flag, and—for retrieved documents—the exact text (raw_text), content_bytes, the IPFS content identifier where applicable (ipfs_cid), the gateway used (archived_via_gateway), and a SHA-256 hash (sha256) of the bytes. Exact bytes and hashes were captured for 70 of the 72 documents; the two not archived are retained with status = "failed" (two HTTP endpoints that returned HTTP 404 at archival time), and their parsed fields remain available in agent_metadata.csv. This file fixes and integrity-stamps the off-chain layer so analyses remain reproducible despite content drift or link rot.

A data dictionary file documents all column names, data types, and units. Descriptive summaries are provided separately in the Data overview.

## Data overview

This section summarizes the descriptive characteristics of the released data to orient users. It reports data properties only and does not test any hypothesis or draw scientific conclusions. The corresponding distributions are collected in Fig. 3.

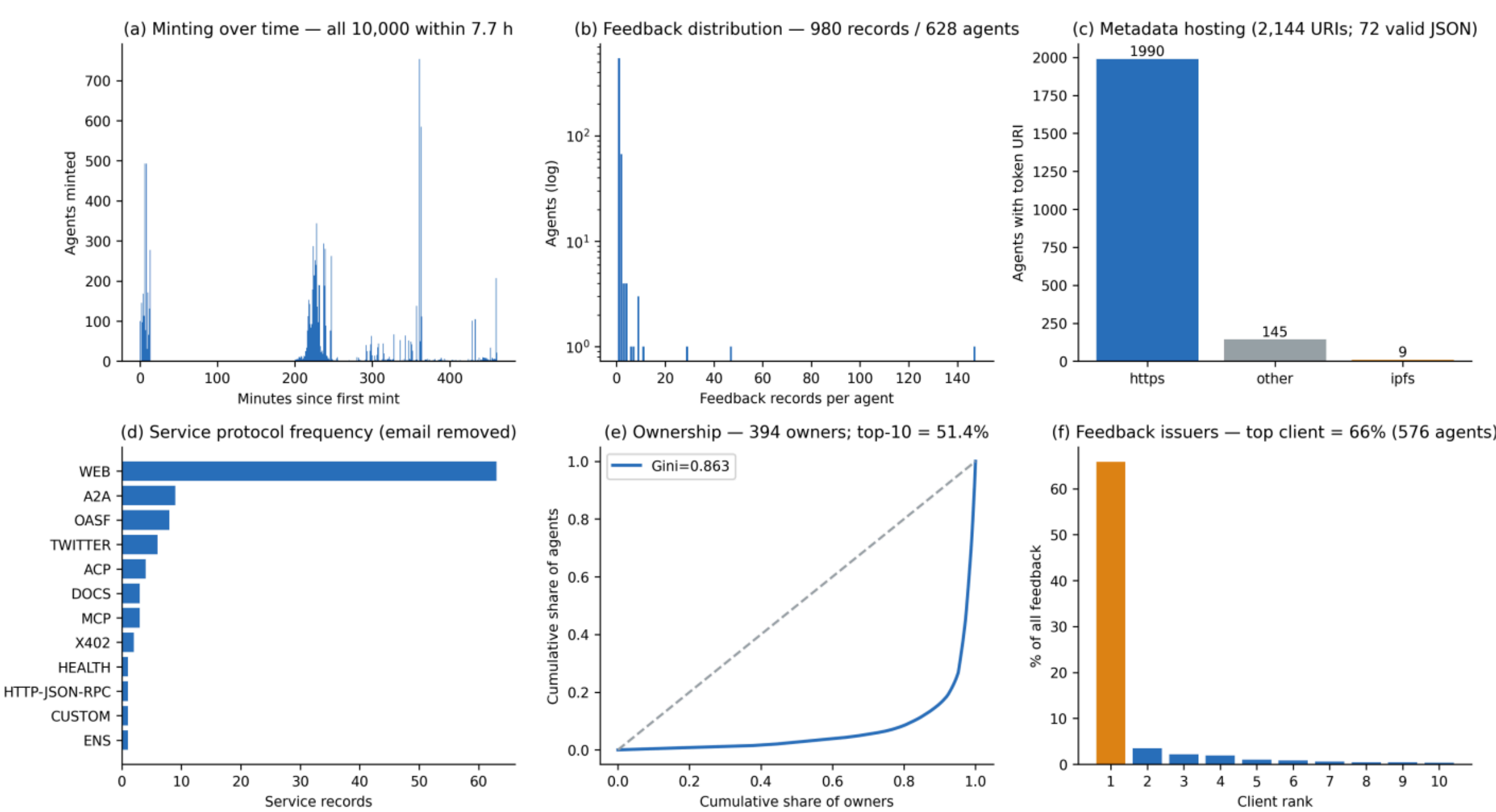

*Figure 3: Data overview. (a) Minting activity over time; (b) feedback records per agent; (c) metadata hosting types among agents with a token URI; (d) declared service-protocol frequency (email endpoints removed); (e) ownership concentration (Lorenz curve); (f) concentration of feedback issuers.*

**Registration timing and batching.** All 10,000 agents were minted on 2026-01-29 within a 7.7-hour interval (blocks 24,339,925–24,342,223). They correspond to 3,028 distinct mint transactions (mean 3.3 agents per transaction; maximum 182), indicating that a substantial share of agents were registered through batch minting (Fig. 3a).

**Identity and metadata coverage.** Of the 10,000 agents, 2,144 carry an on-chain token URI (1,990 HTTPS, 145 other, 9 IPFS); the remaining 7,856 are minted_only. Of the 2,144 token URIs, 72 (3.4%) resolved to valid, parseable JSON metadata, from which 112 service declarations were extracted (Fig. 3c,d). The most common declared service types are general web endpoints, followed by A2A and OASF, with a small number of MCP, x402, and other protocol endpoints.

**Ownership concentration.** The 10,000 agents map to 394 distinct owner wallets; the single largest owner holds 779 agents, and the top ten owners together hold 51.4% of all agents (Gini = 0.863; Fig. 3e), consistent with concentrated, operator-driven batch deployment in this earliest cohort.

**Transfers.** The transfer_history table contains 1,961 ownership-transfer events involving 1,816 agents within the scanned interval; 82.6% occur within the first 10,000 blocks after the first mint.

**Reputation and feedback issuers.** 980 feedback records were submitted for 628 agents by 197 distinct client addresses (Fig. 3b). Feedback issuance is highly concentrated: a single client address submitted 645 records (65.8% of all feedback) to 576 distinct agents, almost entirely under liveness and trust tags, and the top three clients account for 71.4% of all feedback (Fig. 3f). Only 13 clients submitted feedback to more than one agent, and a single record was self-referential (issuer equal to the rated agent's owner). These properties indicate that observable early reputation activity is dominated by a small number of automated monitoring/liveness issuers rather than broad organic peer feedback, which users should account for when interpreting the reputation layer (see Technical Validation).

# Technical Validation

## *Completeness of agent coverage*

The dataset was designed to cover all ERC-8004 agents with identifiers in the range 0–9999. Completeness was verified by comparing the observed agent identifiers against the full expected range. No missing identifiers were detected, indicating that all agents within the target interval were successfully indexed and included in the agents_core table. Transfer activity for ERC-8004 agents was found to be sparse over the observed period, with relatively few ownership changes compared to the total number of registered agents. Reconstructing transfer histories also requires full log traversal rather than direct state queries, incurring substantial computational overhead. We therefore provide transfer data as an auxiliary table derived from a bounded but deliberately optimized scan rather than a full historical reconstruction. The released transfer_history covers the first 100,000 blocks of the observation window (blocks 24,339,925–24,439,925). Because all 10,000 agents were minted within the first 2,298 blocks (~7.7 hours), this interval spans the entire immediate post-registration period for every agent. Within it we recorded 1,961 transfer events involving 1,816 agents; 82.6% occur within the first 10,000 blocks after the first mint, and no transfers were observed in the final ~10,500 blocks of the scanned window (the last event is at block 24,429,384), indicating that ownership activity for this earliest cohort had largely ceased well before the window boundary. This sparsity is intrinsic to the cohort: minting completed within hours and most agents are held by a small number of largely inactive owner wallets (see Data overview). The scanning pipeline—chunked log requests with adaptive sizing, bounded concurrency, rate-limiting, retries, and checkpointing—was designed to extract these sparse events efficiently, and the same methodology can be re-run over later, higher-activity windows to build richer behavioral datasets as the ecosystem matures. Users should therefore interpret transfer_history as a systematically extracted record of early ownership dynamics rather than an exhaustive lifetime ledger; transfers after block 24,439,925 are not captured.

## *Consistency of identity data*

Identity records were derived directly from on-chain events and contract state at a fixed observation block. Minting information, including the block number, transaction hash, and timestamp, was cross-referenced between event logs and block headers. Ownership data and token URIs were retrieved via contract calls at the observation block, ensuring that all identity fields correspond to a consistent on-chain state. As an explicit cross-validation, mint attributes derived from event logs (mint block, transaction hash, timestamp) were reconciled against the corresponding block headers and transaction receipts, and no inconsistencies were found. Agent identifiers were verified to be unique and contiguous over 0–9,999 with no duplicates or gaps.

### *Metadata validation and coverage*

Metadata resolution was performed by retrieving and parsing token-linked URIs. A subset of agents (72 out of 10,000) yielded valid JSON metadata. The primary cause of metadata retrieval failure was that many token URIs resolved to HTTP endpoints returning non-JSON content, preventing structured parsing. This outcome is consistent with the ERC-8004 design, where metadata is optional and externally hosted, and therefore not guaranteed to be uniformly available or machine-readable. Resolution outcomes decompose as follows: 7,856 agents carried no token URI (minted_only) and were not resolvable by construction; of the 2,144 agents carrying a token URI, 72 returned valid JSON, while the remainder failed because the endpoint returned non-JSON content (predominantly HTML landing or error pages), returned an empty body, timed out, or could not be reached. This pattern is consistent with an immature, early-stage ecosystem in which most agents were batch-registered without populating a machine-readable off-chain document. The low coverage therefore reflects ecosystem adoption and hosting practices rather than a sampling choice; it nonetheless limits the metadata and service layers, so analyses depending on declared capabilities are restricted to the 72-agent/112-service subset and should not be generalized to the full population. Moreover, the 72 metadata-bearing agents are a self-selected rather than random subset—operators who chose to publish machine-readable off-chain documents—and therefore skew toward more active, productized projects; the metadata and service layers should be treated as illustrative of early ERC-8004 metadata usage rather than as a representative cross-section of all 10,000 agents.

### *Off-chain reproducibility*

On-chain fields are exactly reproducible because every contract call is pinned to the fixed observation block via historical eth_call; given an archive node, results are provider-independent. To preserve the off-chain layer, the revised release archives the resolved metadata documents in metadata_raw_snapshot.jsonl, capturing the exact bytes and a SHA-256 hash for 70 of the 72 metadata-bearing agents. Integrity is reinforced by source structure. The 72 metadata-bearing agents decompose by source into 3 on-chain data URIs, 10 IPFS documents, and 59 HTTP documents. The 3 data-URI documents are encoded directly in the on-chain token URI and are byte-reproducible from the chain; all 10 IPFS documents are content-addressed; we recorded each one's content identifier (CID), which is itself a cryptographic content hash, so integrity can be verified independently of any gateway. Of these 10, nine use an ipfs:// token URI and one is referenced through an https IPFS-gateway URL. Of the 59 HTTP documents, 57 were retrieved and archived with SHA-256, while 2 returned HTTP 404 at archival time. The two failures are a concrete instance of the link rot that motivates this snapshot; for these two agents, the parsed fields collected at the original resolution remain in agent_metadata.csv. Exact bytes and integrity hashes were thus captured for 70 of the 72 documents (3 + 10 + 57). The snapshot fixes the off-chain content for the large majority of agents and enables integrity verification; for the mutable HTTP subset, the archived bytes reflect the re-resolution date.

### *RPC extraction reliability*

On-chain extraction achieved complete agent-level coverage: identity records were obtained for all 10,000 agents, and reputation summaries for all 10,000 agents. Transient RPC errors were absorbed by exponential-backoff retries, adaptive reduction of the log-scan chunk size on get_logs failures, and a serial second-pass retry. Only 2 of 10,000 agents (0.02%) required additional targeted re-querying for their reputation records; after recovery, no on-chain extraction failures remained. The final unresolved on-chain extraction failure rate is therefore 0% (0 of 10,000 agents).

### *Reputation data integrity*

Reputation data were validated at both the aggregate and record levels. Summary statistics were successfully computed, while individual feedback records were observed for 628 agents, totaling 980 entries. This discrepancy reflects the design of the reputation registry, where summary values may be defined even in the absence of extensive feedback activity.

All feedback entries include raw values and decimal precision, preserving exact on-chain representations. No anomalous values, such as negative feedback scores or invalid decimal ranges, were observed. Additionally, no revoked feedback entries were present in the dataset.

**Reputation integrity and manipulation risk.** ERC-8004 reputation is permissionless—any address may submit feedback—so the layer is, in principle, exposed to Sybil behaviour, collusion, and reputation farming. We do not adjudicate authenticity, but we report issuer-side structure and release a dedicated per-issuer screening table (feedback_issuer_summary.csv) so that users can identify suspicious patterns: for each client address, it gives the feedback count, the number of distinct agents rated, the share of all feedback, a self-referential flag, and a high-volume flag. The structure is highly concentrated: a single automated issuer submitted 645 of 980 records (65.8%; see Data overview), the top three issuers account for 71.4% of records, only 13 issuers rated more than one agent, and one record was self-referential. Users studying trust dynamics should apply issuer-level filtering or clustering and, where appropriate, established blockchain anomaly- and fraud-detection techniques[30] before drawing conclusions from the reputation layer. More broadly, the integrity of the recorded data depends on the security of the underlying smart contracts, and it is useful to distinguish two layers. Our pipeline records canonical on-chain state deterministically at the fixed observation block, and the event-log/header/receipt cross-checks reported under "Consistency of identity data" found no inconsistencies, so the dataset's fidelity to the ledger is independent of contract security. The semantic validity of the underlying records is not: a successful exploit of the identity or reputation registries—e.g., via re-entrancy or access-control flaws enabling unauthorized minting or feedback injection—would write malicious records on-chain that our pipeline would faithfully capture as legitimate, a limitation intrinsic to any on-chain dataset. We therefore flag, as relevant context, systematic smart-contract analysis and vulnerability-detection tooling[14], agent-level threats such as memory poisoning of LLM-based agents[31], and AI-enhanced blockchain-security approaches that aim to detect or mitigate such risks[32–34], and recommend that users requiring assurance of honest provenance combine the issuer-level screening fields provided here with such methods.

### *Reproducibility*

The dataset was constructed using a deterministic pipeline based on a fixed observation block. All contract state queries were executed using historical eth_call, enabling exact reconstruction of the dataset given the same block height and RPC source. The full data collection logic is provided as open-source code, allowing independent verification and regeneration of the dataset. Because all state queries are pinned to a fixed block height and executed via historical eth_call against an Ethereum archive node, the on-chain portion of the dataset is deterministic and provider-independent: any archive node serving the same block height returns identical results. Because the agents documented here are deployed on standard, passive ERC-8004 contracts with a Validation Registry that is not yet populated on-chain, this release also serves as an empirical baseline against which future AI-enhanced deployments — e.g., execution-layer intelligent validation or machine-learning-optimized on-chain indexing—can be compared as the ecosystem matures[32].

## Usage Notes

The dataset is designed for both scientific analysis and practical applications involving ERC-8004 AI agents. The primary table, agents_core, provides a complete index of the first 10,000 registered agents and can be used as an entry point for all downstream analyses.

Researchers may use the dataset to study the emergence of on-chain AI agent identity systems, including the distribution of ownership, the temporal dynamics of agent creation, and the early-stage formation of reputation signals. The availability of both aggregated (agent_reputation_summary) and record-level (agent_feedback_records) data enables analysis of reputation mechanisms, including feedback sparsity, client participation, and the use of multi-dimensional tags such as liveness, trust, and quality.

The metadata and service tables (agent_metadata and agent_services) provide a partial but structured view of how agents expose capabilities and endpoints. These tables can be used to analyze interoperability patterns across emerging protocols such as MCP, A2A, OASF, and x402, as well as to build registries or discovery systems for agent services.

Because metadata is externally hosted and optional, users should account for incomplete coverage when performing analyses that depend on descriptive fields or service declarations. Similarly, feedback data is sparse relative to the full agent population, reflecting the early-stage adoption of the reputation registry. Note also that direct contact endpoints (email addresses), which appeared in a small number of off-chain service declarations, have been removed from the released agent_services table to avoid distributing personal-style contact identifiers; all other declared endpoints (e.g., web, docs, social) are retained, and the original off-chain documents remain publicly retrievable via each agent's on-chain token URI. Because the resolved metadata come from a self-selected subset of operators, the metadata and service tables are illustrative rather than representative of the full agent population.

All numerical values in the reputation layer are stored in raw integer form together with decimal precision. Users should convert these values appropriately depending on their analytical requirements.

The dataset is anchored to a fixed observation block. Users interested in longitudinal dynamics or agent lifecycle evolution should reconstruct additional snapshots at different block heights using the provided codebase.

## Data Availability

The dataset described in this study is publicly available via the Harvard Dataverse repository. It includes all tables described in the Data Records section, including agents_core, agent_metadata, agent_services, agent_reputation_summary, agent_feedback_records, mint_economics, transfer_history, crosschain_registrations, feedback_issuer_summary, metadata_raw_snapshot.jsonl and a data dictionary file.

The dataset is provided in machine-readable CSV format to facilitate reuse and integration into downstream analysis pipelines. The version reported in this study corresponds to a snapshot of Ethereum mainnet at block 24,839,925.

A persistent identifier (DOI) is assigned to the dataset via Harvard Dataverse to ensure long-term accessibility and reproducibility. The dataset[29] is available at: https://doi.org/10.7910/DVN/HJZW8Q.

The dataset is released under a CC0 1.0 public-domain dedication. It contains factual on-chain records and machine-readable functional metadata fields with URL pointers only; no externally hosted media (images, audio, video, or full web pages) is redistributed as part of this dataset.

## Code Availability

All code used to collect, process, and validate the dataset is publicly available on GitHub. The repository includes scripts for data collection from Ethereum using Web3.py, metadata resolution, reputation extraction, and database construction.

The codebase is organized as a reproducible pipeline with configurable parameters, including RPC endpoint, block range, and database connection settings. It supports re-execution to reconstruct the dataset from the same observation block.

The code is available at: https://github.com/YulinLiu20/ERC8004

## Funding

This work was supported by Haute Intelligence AG. The funder had no role in the study design, data collection, data analysis, data interpretation, or the preparation of the manuscript, and is not associated with any ERC-8004 agent, contract, or address in the dataset; all data derive from public Ethereum mainnet and public off-chain sources. The research was conducted independently by the author.

## Acknowledgement

The author would like to thank Dylan Figueiredo, Julien Aerni, Siméon Fluck, and all the attendees of the Swiss QuantEcon AI Workshop 2026 for their valuable feedback and comments.

## Competing interests

The author's institution has a consulting agreement with Haute Intelligence AG, which also funded this research (see Funding). Neither the funder nor the contracting party is a stakeholder in the ERC-8004 protocol or in any agent, contract, or address represented in this dataset. The research questions, study design, data collection, analysis, and conclusions were determined independently by the author, and neither the funder nor the contracting party had any role in these or in the decision to publish. The author declares no other competing interests.